# COB-2023-0803
# FLUID DYNAMIC SIMULATIONS OF MACH AND REGULAR REFLECTIONS IN OBLIQUE SHOCK-WAVE CONFIGURATIONS USING ADAPTIVE MESH REFINEMENT

**Sebastian Valencia, Cesar Celis**
Mechanical Engineering Section, Pontificia Universidad Católica del Perú
Av. Universitaria 1801, San Miguel, 15088, Lima, Peru
svalenciar@pucp.edu.pe, ccelis@pucp.edu.pe

**Andrés Mendiburu Zevallos**
International Research Group for Energy Sustainability (IRGES)
Universidade Federal do Rio Grande do Sul
Rua Sarmento Leite 425, Porto Alegre, RS, Brazil
andresmendiburu@ufrgs.br

**Luis Bravo**
DEVCOM - US Army Research Laboratory
Aberdeen Proving Ground, MD 21005
luis.g.bravorobles.civ@army.mil

**Prashant Khare**
Department of Aerospace Engineering
University of Cincinnati
Cincinnati, OH 45221-0070
Prashant.Khare@uc.edu

*Abstract. In the context of the interaction between a moving plane shock wave and an inclined wall (wedge), it is possible to distinguish four distinct shock reflection configurations. These shock wave reflections, which depend on the characteristics of the incident shock wave and the geometry of the surface that it interacts with, are (i) regular reflection (RR), (ii) simple Mach reflection (SMR), (iii) transition Mach reflection (TMR), and (iv) double Mach reflection (DMR). The impact of these shock reflections on flow properties can be significant so understanding them is important when predicting the behavior of shock waves in more complex flow configurations. Previous research works have explored the referred shock reflections through both numerical and experimental approaches, employing various gases and different flow and geometrical configurations. The present study involves the use of a high-fidelity computational fluid dynamics (CFD) tool, known as PeleC, which is a compressible solver based on AMReX specifically designed to handle complex flow configurations. Accordingly, by solving the time-dependent Euler equations for various 2D flow configurations, this work studies shock wave reflections accounting for four different Mach-based operating conditions and compares and analyzes the resulting density profiles on the wedge wall with experimental data. To strike a balance between model accuracy and computational efficiency, adaptive mesh refinement (AMR) is incorporated, and a mesh independence study is performed by varying the number of AMR levels. The numerical method utilized here is based on a finite volume discretization, involving approximate Riemann solvers. Temporal and spatial integration is performed using the method of lines (MOL), a second-order characteristic-based spatial method, coupled with a Runge-Kutta time integration. The time step obeys a specified Courant-Friedrichs-Lewy (CFL) condition of 0.3. The results of this study demonstrate the capabilities of the CFD tool employed as it accurately predicts the sensitivity of wave characteristics to different operating conditions. The findings of this work will serve as a foundation for future studies involving more complex flow configurations such as those featuring detonation waves.*

## 1. INTRODUCTION

Shock wave reflection is a fundamental phenomenon in gas dynamics that has attracted the attention of researchers for over a century. It occurs when a shock wave encounters a surface or another shock wave. The phenomenon was first observed by Ernst Mach in 1878, who experimentally identified two types of reflection configurations, (i) regular reflection (RR) and (ii) Mach reflection (MR) (Mach, 1878). In RR, an incident shock wave and a reflected shock one meet at a reflection point on a reflecting surface. A MR involves in turn one slipstream and three shock waves, the incident, the reflected and the Mach stem ones. Later on, von Neumann proposed the two- and three-shock theories for treating RR and MR, respectively, assuming the flow of an ideal gas to be inviscid (von Neumann, 1943). White and Smith later identified four shock reflection patterns, (i) RR, (ii) single-Mach reflection (SMR), (iii) transitional-Mach reflection (TMR), and (iv) double-Mach reflection (DMR) (White, 1952; Smith, 1945). In SMR, the point of convergence between the incident and reflected shock waves is situated above the wedge. At this convergence point, a third shock known as the Mach stem extends towards the surface of the wedge. Additionally, a curved shear layer called the slipstream trails behind the triple shock convergence point as the shocks propagate along the wedge. In DMR, a bend occurs in the reflected shock wave, giving rise to a second Mach stem. TMR marks the onset of double Mach reflection, where the second triple point is barely visible and manifests itself as a slight bend in the reflected shock (Hryniewicki et al., 2016). The detachment criterion between RR and MR was further investigated by Henderson and Lozzi (1975), who established that, if the transition follows the detachment criterion, a discontinuous transition pressure jump exists. Hence, another criterion named the mechanical equilibrium one is defined based on flow configurations that always fulfil the mechanical equilibrium during transition processes between Mach and regular reflections. These criteria resulted in transitional boundaries that distinguish the different shock wave reflection regions (Henderson and Lozzi, 1975). Deschambault and Glass (1983) conducted experimental investigations that demonstrated that, when compared to the ideal gas case, the use of properties of real gases do not significantly affect the four types of shock wave reflections. They obtained indeed reliable data for the four shock reflection types in air by utilizing infinite-fringe interferometric techniques (Deschambault & Glass, 1983). Ben-Dor et al. (1977) studied in turn a planar shock reflection over a plane double wedge and considered several complicated wave configurations (Ben-Dor et al., 1987). Further studies on the reflection of planar shock waves over different solid walls have been performed in the past both numerically and experimentally (Previtali et al. 2015; Zhang et al., 2016; Geva et al., 2017; Hryniewicki et al., 2017). Accordingly, to assess the capabilities of the computational tool employed here, this work carries out a detailed analysis of regular and Mach reflections using PeleC, a compressible solver based on AMReX. The numerical results obtained in this work are compared with the experimental data from Deschambault and Glass (1983) showing that the numerical model employed here is capable of accurately capturing the supersonic flow characteristics of regular and Mach reflections.

## 2. THEORETICAL ANALYSIS

### 2.1. Principles of Regular Reflection

Normal shock waves are characterized by a sudden and significant increase in pressure and temperature, as well as a decrease in velocity, across the shock wave. The properties of normal shock waves can be calculated using the Rankine-Hugoniot relations (Houghton, 2017), which relate the upstream and downstream states of the gas to the properties of the shock wave. These relations describe the conservation of mass, momentum, and energy across the shock wave. The properties of normal shock waves depend on the Mach number of the incoming flow $M_i$. Based on the conventional Rankine–Hugoniot equations, the detailed flow characteristics are computed as follows (Houghton, 2017),

$$\frac{\rho_2}{\rho_1} = \frac{(1+\gamma)M_i^2}{(\gamma-1)M_i^2 + 2} \tag{1}$$

$$\frac{P_2}{P_1} = \frac{2\gamma M_i^2 - (\gamma-1)}{\gamma+1} \tag{2}$$

$$\frac{a_2}{a_1} = \sqrt{\frac{T_2}{T_1}} = \sqrt{\frac{P_2 \rho_1}{P_1 \rho_2}} \tag{3}$$

$$\frac{U_2 - U_1}{a_1} = \frac{2(M_i^2 - 1)}{M_i(\gamma+1)} \tag{4}$$

where the flow properties temperature $T$, density $\rho$, pressure $P$, sound speed $a$, and velocity $U$ denoted by the subscript 1 characterize the flow conditions before the shock, while those properties denoted by subscript 2 correspond to the flow conditions after the shock. $\gamma$ is in turn the specific heat ratio.

### 2.2. Analytical Boundaries for RR-MR transition

The theoretical boundaries that define the RR-MR transition, i.e., the detachment boundary, the sonic boundary, and the von Neumann boundary, are obtained from the analytical solutions determined by von Neumann (1943). The work of

von Neumann was later studied by Henderson and Lozzi (1975), who provided an analytical expression for these transition boundaries. Figure **1** shows that there are three regions determined by the mechanical equilibrium and detachment boundaries, (i) an upper region only for regular reflection, (ii) a dual region for either regular reflection or Mach reflection between the two boundaries, and (iii) a lower region for only Mach reflection (SMR, TMR, and DMR).

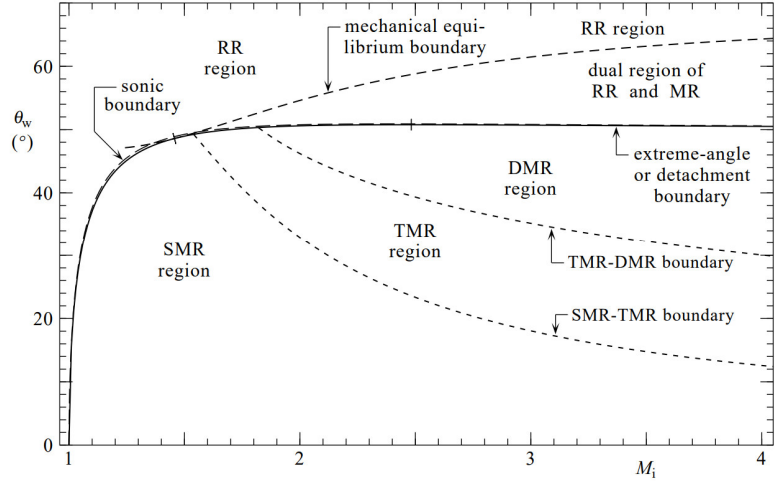

Figure 1. Regions of RR and MR patterns separated by analytical transition boundaries for air (Hryniewicki et al., 2016).

The physically realizable solution that defines the detachment boundary comes from the two-shock theory (von Neuman, 1943). Expressed in terms of wedge angle $\theta$ and incoming flow Mach number $M_i$, this solution is given by,

$$\cos\theta = \frac{1}{a + 2e\cos(f/3)}. \tag{5}$$

where,

$$a = \frac{1 + (\gamma - 1)d}{3}, \quad b = 2d - d^3, \quad c = \gamma d^2, \quad d = \frac{2}{\gamma + 1}\frac{M_i^2 - 1}{M_i^2}, \quad e = \sqrt{a^2 + \frac{b}{3}},$$

$$f = \cos^{-1}\left(\frac{ab + 2a^3 - c}{2e^3}\right), \tag{6}$$

The mechanical equilibrium criterion in turn is established based on the three-shock theory (von Neuman, 1943). The transition from Mach reflection to regular reflection appears when the triple-point trajectory angle diminishes to zero. At this situation, the Mach stem decreases to an infinitesimal length and the slipstream disappears. The relation between the wedge angle $\theta$ and incoming flow Mach number $M_i$ is shown to be,

$$\cos^2\theta = \frac{c}{b + \sqrt{b^2 - ac}} \tag{7}$$

where,

$$a = 4d + 2(\gamma - 1)(\gamma + 2)d^2 - (\gamma^2 - 1)d^3, \quad b = \gamma + 3 - \frac{1}{2}(5 - \gamma)(\gamma + 1)d + 2\gamma d^2,$$

$$c = 4 - 4d, \quad d = \frac{2}{\gamma + 1}\frac{M_i^2 - 1}{M_i^2}. \tag{8}$$

Finally, the sonic boundary is given by a fifth order polynomial in terms of $\sin^2\theta$ versus the inverse incident shock strength $P_1/P_2$, and it is not considered here because this boundary lies very close to the detachment one, differing by less than a half of a degree for each given value.

## 3. MATHEMATICAL MODELING

Accounting for two-dimensional flow configurations, the time-dependent Euler equations are solved here. This choice of using the Euler equations aligns with previous studies conducted by various researchers in the field. Therefore, for a 2D Cartesian coordinate system (x, y), the corresponding transport equations for mass, momentum, and energy expressed in matrix-vector form are given by (Houghton, 2017),

$$\frac{\partial \boldsymbol{U}}{\partial t} + \frac{\partial \boldsymbol{F}}{\partial x} + \frac{\partial \boldsymbol{G}}{\partial y} = 0, \tag{9}$$

where $t$ stands for physical time. The vector of solution variables, $\boldsymbol{U}$, and the inviscid flux vectors, $\boldsymbol{F}$ and $\boldsymbol{G}$, are in turn given by,

$$\boldsymbol{U} = \begin{pmatrix} \rho \\ \rho u \\ \rho v \\ \rho E \end{pmatrix}, \qquad \boldsymbol{F} = \begin{pmatrix} \rho u \\ \rho u^2 + p \\ \rho uv \\ (\rho E + p)u \end{pmatrix}, \qquad \boldsymbol{G} = \begin{pmatrix} \rho v \\ \rho uv \\ \rho v^2 + p \\ (\rho E + p)v \end{pmatrix}, \tag{10}$$

where $\rho$, $u$, $v$, and $p$ represent gas density, velocities along the x and y coordinates, and pressure, respectively. In addition, $E$ is the total energy including the kinetic ($E_k$) and internal ($E_U$) energies,

$$E = E_k + E_U, \tag{11}$$

$$E_k = \frac{1}{2}(u^2 + v^2), \tag{12}$$

$$E_U = \frac{R}{\gamma - 1}T, \tag{13}$$

where $T$ and $R$ stand for, respectively, flow temperature and gas constant.

## 4. NUMERICAL MODELING

This section highlights the numerical modeling approach utilized here, with specific focus on the solver and numerical schemes employed, the geometric configuration accounted for, and the boundary conditions imposed.

### 4.1. Solver and numerical schemes

To conduct the intended numerical simulations, an open-source AMR-based compressible reacting flow solver, named PeleC (PeleC, 2023), is employed in this work. PeleC solves transport equations for mass and momentum in the compressible flow regime. PeleC is built on top of the AMReX framework, which massively facilitates parallel block-structured adaptive mesh refinement (AMR). The validity of PeleC has been established through its previous successful applications in several standard cases, including the Sod shock tube (Henry de Frahan et al., 2022). The governing equations here are closed with the ideal gas equation of state (EoS) available in the PelePhysics submodule, which provides models and parameters associated with thermodynamics, transport properties, and chemical reactions.

The system of partial differential equations solved here is spatially discretized using a second-order finite volume approach. Notice that PeleC supports two different discretization methods, (i) the unsplit piecewise parabolic method (PPM) with optional hybrid PPM WENO variants, and (ii) a second-order characteristic-based spatial method coupled with a Runge-Kutta time integration known as a method of lines (MOL). For the present study, only the former method has been utilized as it is suitable for complex geometries with embedded boundaries (EB). In addition, for an accurate resolution of the resolving shock waves, adaptive mesh refinement (AMR) is enabled at locations with relatively high-density gradients.

### 4.2. Geometric configuration

A two-dimensional computational domain has been adopted here which shares similarities with the one used by Hryniewicki et al. (2016), Figure **2**. More specifically, the computational domain spans over a spatial region with x-coordinates ranging from 0 to 4 meters, and y-coordinates ranging from 0 to 0.75 meters. The domain is discretized using a grid featuring 512 cells in the $x-$direction and 96 cells in the $y-$direction. It is worth noticing that, in the two-dimensional plane, the grid cells are square shaped with an aspect ratio of unity ($Dx/Dy = 1$). Besides, the initial position of the shockwave has been established at $x = 0.5$ meters, whereas the rigid wedge is introduced at $x = 3.0$ meters with different wedge angles. To enhance the grid resolution in specific regions of the flow, adaptive mesh refinement (AMR) processes are carried out during the computations. Both to reduce numerical errors and to ensure that grid-independent results are obtained, a mesh independence study was also conducted. The referred study, which main outcomes are summarized in Section 5.1, involved analyzing the results obtained using different AMR levels.

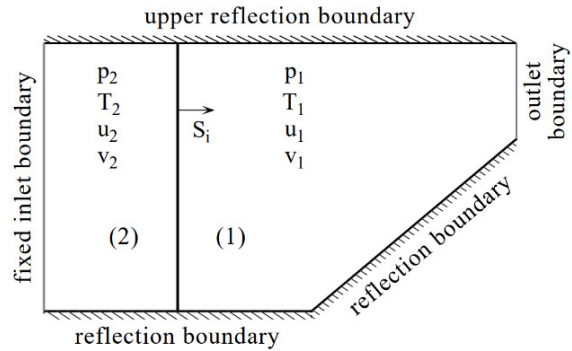

Figure 2. 2D domain for shock wave reflections of an inclined and rigid wedge with angle $\boldsymbol{\theta}$ (Hryniewicki et al., 2016).

**4.3. Boundary and initial conditions**

It is of particular interest here to analyze the reflected shock waves originated under different incoming flow Mach numbers and wedge angles. To do so, for each of the four cases studied here and listed in Table 1, the initial thermochemical conditions ahead of the incident shock wave (Figure 2, region 1) are identical to those investigated by Deschambault and Glass (1983). These initial conditions include density, temperature, and pressure of the fluid flow. Furthermore, the flow properties behind the incident shock (Figure 2, region 2) are determined using the Rankine-Hugoniot equations, Eqs. (1)-(4).

Table 1. Initial conditions for the four oblique shock wave reflection cases studied.

| CASE | $\boldsymbol{\theta_w}$ | $\boldsymbol{M_s}$ | $\boldsymbol{P_1(Kpa)}$ | $\boldsymbol{T_1(K)}$ | $\boldsymbol{\rho_1\left(\frac{kg}{m^3}\right)}$ |
|---|---|---|---|---|---|
| 1 | 63.4° | 2.05 | 33.3306 | 298.4 | 0.387 |
| 2 | 60° | 4.70 | 6.13283 | 298.5 | 0.0712 |
| 3 | 27° | 2.03 | 33.3306 | 299.2 | 0.387 |
| 4 | 20° | 7.19 | 7.99934 | 298.5 | 0.0929 |

Regarding the boundary conditions, all numerical simulations carried out in this work involved the use of a first-order extrapolation (FOExtrap) based outflow boundary condition along the y-direction boundaries. In addition, the upper, lower, and wedge boundaries were set to no-slip wall conditions. Finally, notice that, as shown in Figure 3, each of the four cases analyzed in this work correspond to a specific shock wave reflection region defined by the detachment and mechanical equilibrium criteria.

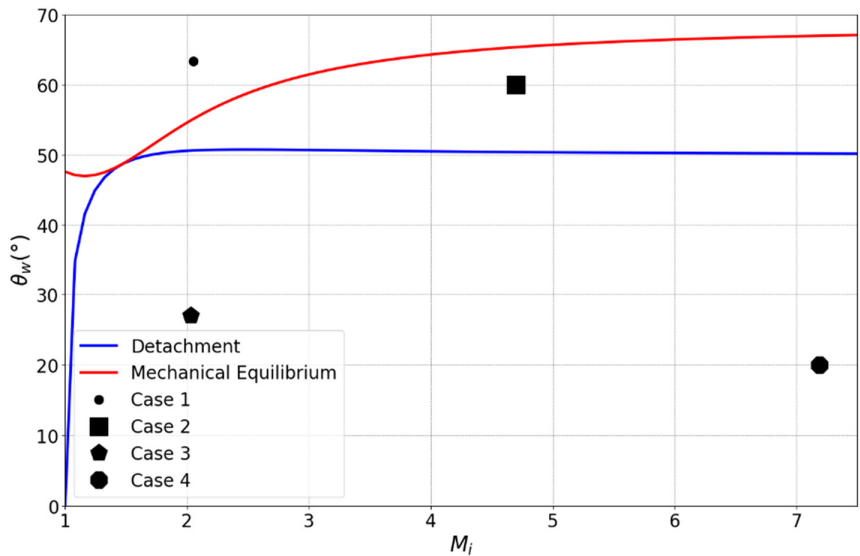

Figure 3. Oblique shock wave reflection cases studied and their relative location regarding the analytical transition boundaries.

## 5. RESULTS AND DISCUSSION

The main numerical results obtained in this work are presented and discussed in this section. Both qualitative and quantitative analyses of the referred results are carried out.

### 5.1. Mesh independence study

A mesh independence analysis was firstly performed here to determine the requirements in terms of grid resolution for the numerical simulations. Initially, a mesh with 512x96 elements and a minimum element size of 7.81 mm was generated and used as the base mesh. On top of this initial mesh, four other meshes were generated by varying the number of AMR levels. Finally, accounting for Case 3 (Table 1), several numerical simulations were conducted using the five meshes generated and the corresponding results are shown in Figure **4**. As illustrated in Figure **4** (left plot), in terms of wedge wall density, there are no significant differences between the results obtained with the meshes featuring 3 and 4 AMR levels. Consequently, the mesh including 3 AMR levels, featuring a minimum element size of 0.97 mm and a mesh size of about 1.5 million elements, was chosen here to carry out the intended numerical simulations. This mesh configuration was deemed adequate to achieve the desired level of accuracy in the simulations performed in this work.

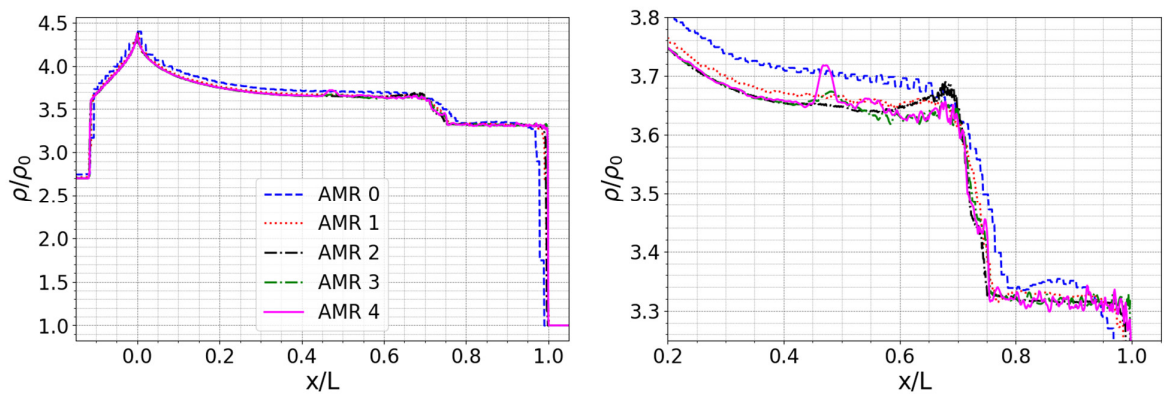

Figure 4. Density ratio profiles along the compression ramp surface obtained with meshes featuring different number of AMR levels. Right plot is a zoom of the left one.

Figure **5** shows the computational mesh utilized here for the numerical simulations carried out. In this figure, the black and gray boxes indicate the regions where adaptive mesh refinement (AMR) was employed. More specifically, the black boxes represent the areas with the highest level of refinement, whereas the gray ones indicate regions with a lower level of refinement. It can be observed from Figure **5** that the area around the wedge surface features the highest level of refinement throughout. This occurs because this wedge region is a critical zone where the shock wave interacts with the wedge wall. As such, it is essential to have a refined and high-quality mesh to both capture the intricate features of the flow and accurately predict the wave-wall reflection interactions. Overall, the mesh was designed to ensure the accuracy of the numerical results by enforcing sufficient mesh resolution in the regions of interest.

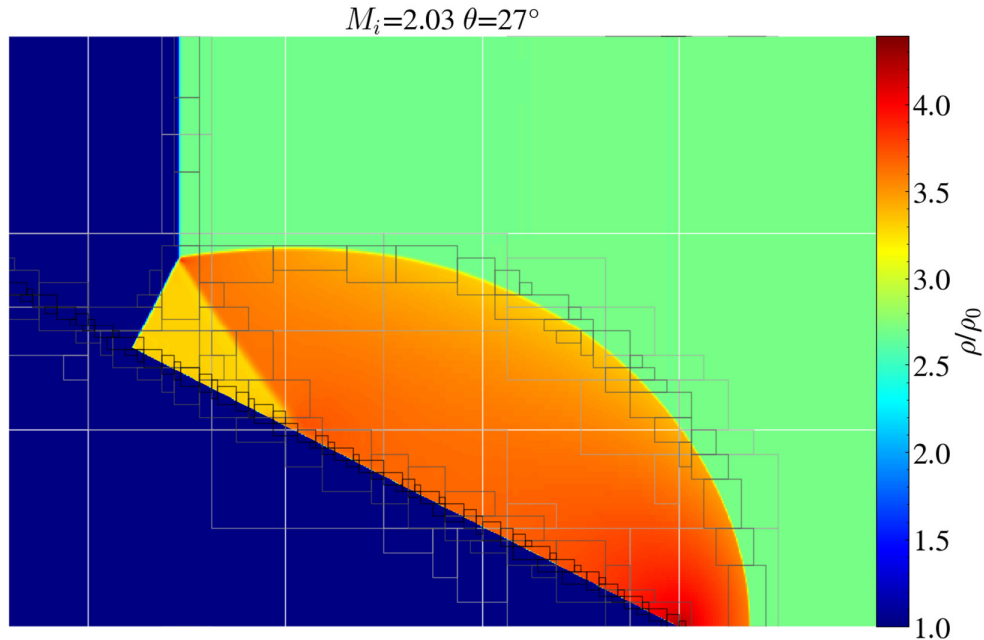

Figure 5. Details of computational mesh employed plus AMR levels included and contours of density ratio.

**5.2. Density ratio distributions**

Figure **6** to Figure **9** shows density ratio contours obtained from the numerical simulations conducted for the four cases studied here (Table 1), which allows a qualitative comparison with the experimental data presented by Deschambault and Glass (1983). To gain an insight of the observed patterns, it is imperative to delve into the theories and descriptions of regular reflection, single Mach reflection, and transitional Mach reflection. Regular reflection occurs when a shock wave encounters a solid wall at an appropriate angle. In this scenario, the resulting reflected shock wave remains attached to the wall, forming an oblique shock wave. Besides, the incident and reflected shock waves intersect, giving rise to a regular arrangement of shock waves. As supported by the revised transition boundaries theory depicted in Figure 3, Case 1 and Case 2 studied here belong to the region of regular reflection. Both the numerical results and the experimental data corroborate this finding, as no Mach stem is present, and the shock pattern includes only two propagating shock waves (Figure **6** and Figure **7**). In single Mach reflection in turn, the incident shock wave strikes the surface of the wedge, generating a curved reflected shock wave that intersects with the incident one. The point of intersection forms what is called the triple point, which lies above the surface of the wedge. At the triple point, a third shock wave called the Mach stem extends towards the surface of the wedge. This shock pattern including the triple point is clearly noticed in the numerical results obtained for Case 3 and the experimental data from Deschambault and Glass (1983) (Figure **8**). Finally, transitional Mach reflection involves a second triple point that represents the confluence point of the incident and reflected shocks. This second triple point is slightly visible and manifest itself as a subtle kink or bend in the reflected shock wave. In this scenario, the overall shock pattern is in the process of transitioning from the simpler single Mach reflection pattern to the more intricate double Mach reflection one. Case 4 (Figure **9**) studied here exemplifies this scenario, where the second triple point is barely discernible. Both the numerical results and the experimental data exhibit this trend, which agree with the relevant theory and the von Neumann criteria.

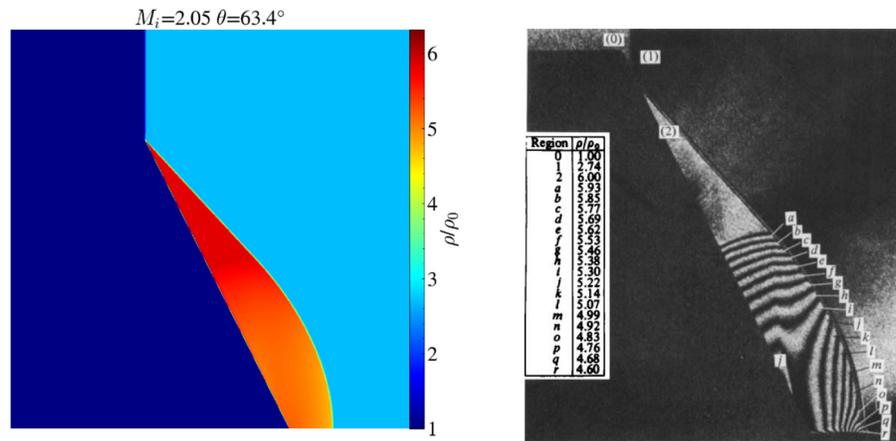

Figure 6. Density ratio contours for Case 1 compared with experimental data from Deschambault and Glass (1983).

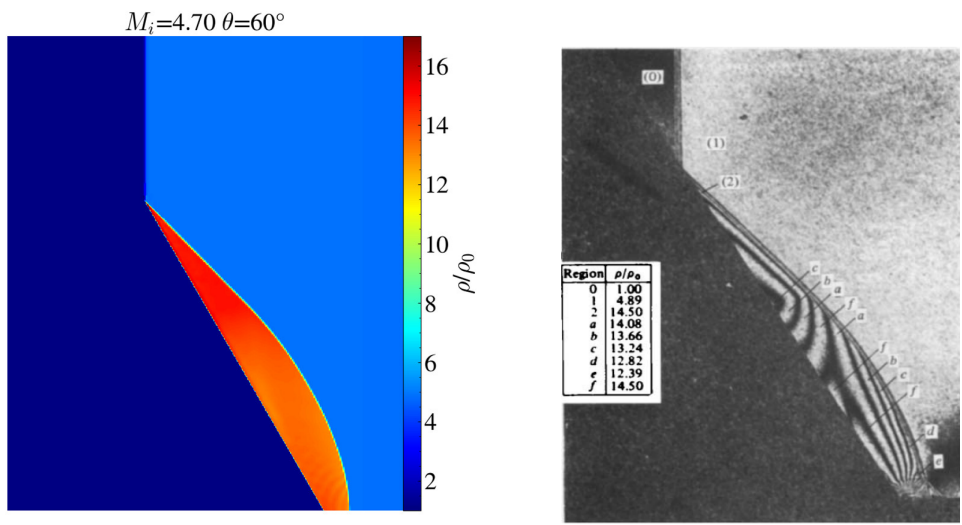

Figure 7. Density ratio contours for Case 2 compared with experimental data from Deschambault and Glass (1983).

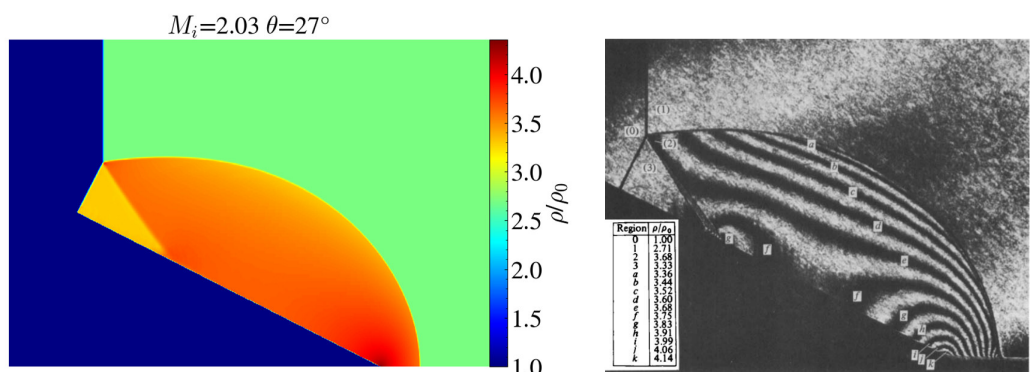

Figure 8. Density ratio contours for Case 3 compared with experimental data from Deschambault and Glass (1983).

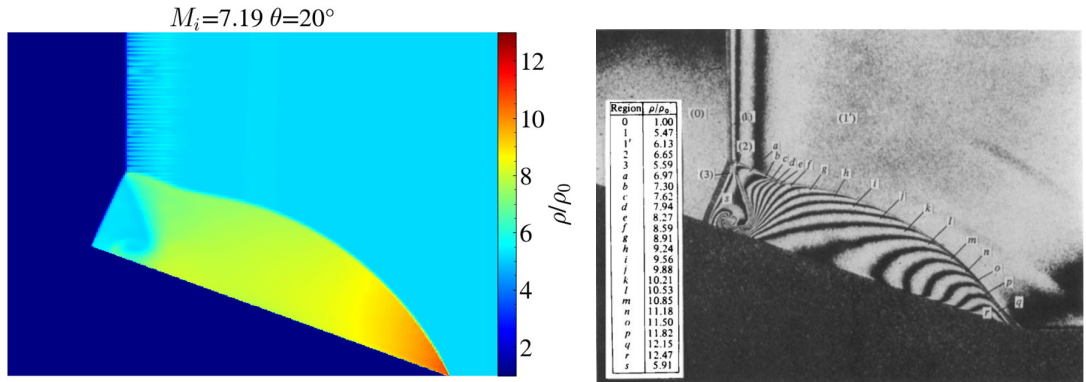

Figure 9. Density ratio contours for Case 4 compared with experimental data from Deschambault and Glass (1983).

### 5.3. Density ratio profiles

Figure **10** illustrates with blue lines the density ratio profiles along the wedge wall computed for each of the four cases studied here (Table 1). This figure also includes the experimental data obtained by Deschambault and Glass (1983) as red symbols. From Figure **10**a, for Case 1, featuring a Mach number of 2.05 and an angle of incidence of 63.4°, and exhibiting a regular reflection, the numerical results show a relatively good agreement with the experimental data, with no major differences detected in this case. This emphasizes that the numerical predictions accurately captured the supersonic flow characteristics. In Case 2 (Figure **10**b), which features a Mach number of 4.70 and an angle of incidence of 60°, the flow is expected to be in the dual region of Regular and Mach reflections. In this case, the numerical results show a pattern similar to a regular reflection profile in the wedge wall, which is not observed in the experiments. This discrepancy could be attributed to the limitations of the numerical model employed in this work, as it may not be able to fully capture the complex physics of the flow in this particular situation. Nevertheless, the results of the numerical simulations carried out for this case still provide valuable insights into the flow characteristics and indicate the need for further research and development of more accurate models. In Case 3 (Figure **10**c), characterized by a Mach number of 2.03 and an angle of incidence of 27°, a SMR is obtained (Figure **8**). Like Case 1, in this case the numerical results are quite similar to the experimental data and the discrepancies are insignificant. Finally, in Case 4 (Figure **10**d), featuring a Mach number of 7.19 and an angle of incidence of 20°, a TMR is obtained. The numerical results and the experimental data show in this case some discrepancies in the region near the triple point and at the reflected shock. However, the density profile is well captured along all the wedge wall, providing valuable insights into the shockwave interactions and supersonic flow characteristics.

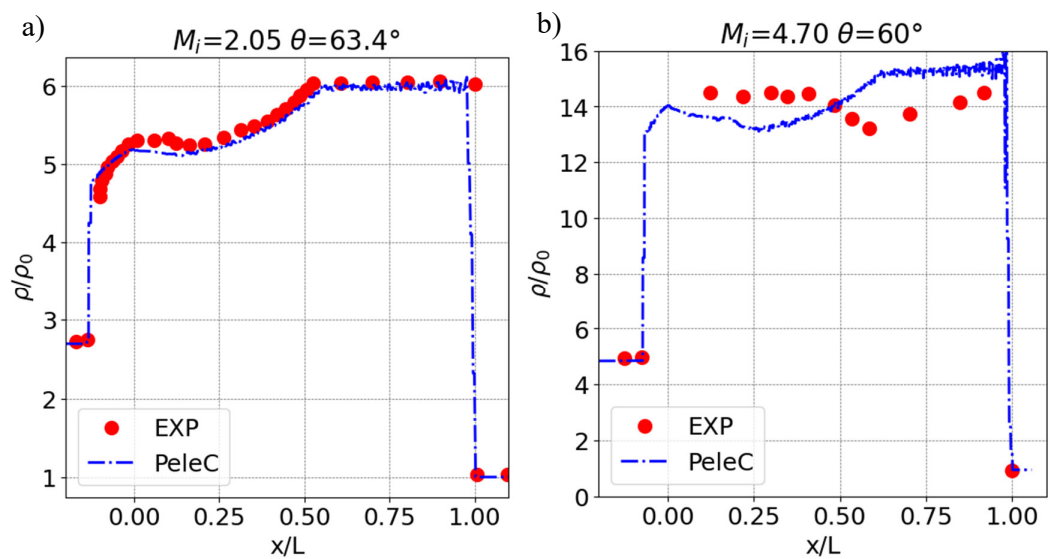

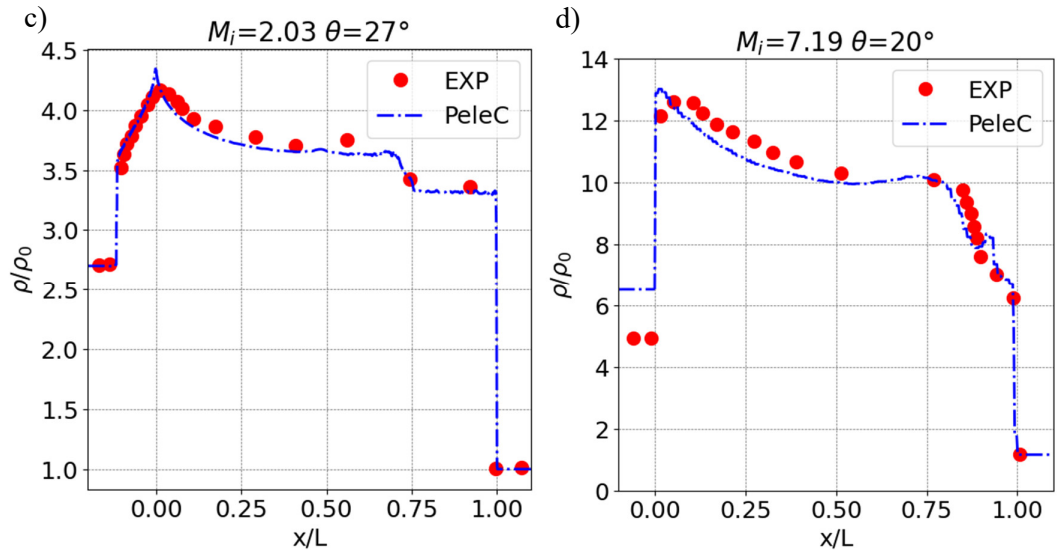

Figure 10. Density ratio profiles along the compression ramp surface for different Mach numbers and wedge angles. Blue lines: numerical results. Red symbols: experimental data from Deschambault and Glass (1983).

## 6. CONCLUSIONS

This study was particularly focused on regular (RR) and Mach (MR) reflections, which are the two main types of shock wave reflections observed in steady flows. The computational tool PeleC, which includes adaptive mesh refinement (AMR), was used to model the shock wave reflection phenomena, and the obtained numerical results were compared with experimental data available in literature. From the results obtained in this work, it can be concluded that the numerical model employed here is able to accurately capture the supersonic flow characteristics for cases such as those involving regular and single Mach reflections. However, for cases involving more complex physics, such as those featuring dual regions of regular and Mach reflections, the model seems to have limitations that affect its accuracy. Overall, the results discussed here provide valuable insights into shock wave interactions and characteristics of wedge flows. The particular supersonic flow configurations where the numerical model seems to have some limitations highlight the need for further development and use of more accurate models. Notice that the findings of this work contribute to ongoing research in the field of supersonic flows focused on the study of more complex fluid flows such as those featuring 3D configurations, viscous flows, and detonation waves.

## 7. ACKNOWLEDGEMENTS

This work has been supported by the US Army Research Laboratory under Research Grant No. W911NF-22-1-0275. Luis Bravo was supported by the US Army Research Laboratory 6.1 Basic research program in propulsion sciences.

## 9. RESPONSIBILITY NOTICE